\documentclass[11pt]{article}

\usepackage[utf8]{inputenc}
\usepackage[english]{babel}
\usepackage{amsmath}
\usepackage{amsfonts}
\usepackage{amssymb}
\usepackage{ifpdf}

\ifpdf
\usepackage[T1]{fontenc}
\usepackage[pdftex]{graphicx}
\else
\usepackage{graphicx}
\fi 

\usepackage{authblk}

\usepackage{mathtools,xfrac}
\usepackage[pdfpagelabels=false,colorlinks=true,allcolors=black]{hyperref}

\def\bit  {\begin{itemize}}
\def\eit  {\end{itemize}}
\def\beq  {\begin{equation}}
\def\eeq  {\end{equation}}
\def\beqa {\begin{eqnarray}}
\def\eeqa {\end{eqnarray}}
\def\bef  {\begin{figure}}
\def\ef   {\end{figure}}

\def\ben  {\begin{enumerate}}
\def\een  {\end{enumerate}}
\def\bec  {\begin{center}}
\def\ec   {\end{center}}
\def\bet  {\begin{tabbing}}
\def\et   {\end{tabbing}}
\def\vs   {\vspace{0.5cm}}

\newcommand \be  {\begin{equation}}
\newcommand \bea {\begin{eqnarray} \nonumber }
\newcommand \ee  {\end{equation}}
\newcommand \eea {\end{eqnarray}}

\title{{\bf Application of spin glass ideas in social sciences, \\economics and finance} \vs\\
Contribution to the edited volume \\"Spin Glass Theory \& Far Beyond - Replica Symmetry Breaking after 40 Years", World Scientific, 2023 (to appear).
\vs}

\author[1,2,3]{Jean-Philippe Bouchaud}
\author[4]{Matteo Marsili}
\author[5,6]{Jean-Pierre Nadal}
\affil[1]{Chair of Econophysics and Complex Systems, \'Ecole polytechnique,\\ 91128 Palaiseau Cedex, France }
\affil[2]{Capital Fund Management, 23 Rue de l’Universit\'e, 75007 Paris, France}
\affil[3]{Acad\'emie des Sciences, 23 Quai de Conti, 75006 Paris, France}
\affil[4]{The Abdus Salam International Centre for Theoretical Physics, 34151 Trieste, Italy}
\affil[5]{Laboratoire de Physique de l’Ecole Normale Sup\'erieure, ENS, Universit\'e PSL, CNRS, Sorbonne Universit\'e, Universit\'e Paris Cit\'e, 24 rue Lhomond, F-75005 Paris, France}
\affil[6]{Centre d'analyse et de math\'ematique sociales, EHESS, CNRS, Ecole des Hautes Etudes en Sciences Sociales, 54 bd Raspail, F-75006 Paris, France}

\date{March 20, 2023}

\begin{document}
\maketitle

Classical economics has developed an arsenal of methods, based on the idea of representative agents, to come up with precise numbers for next year's GDP, inflation and exchange rates, among (many) other things. Few, however, will disagree with the fact that the economy is a complex system, with a large number of strongly heterogeneous, interacting units of different types (firms, banks, households, public institutions) and different sizes.   

Now, the main issue in economics is precisely the emergent organization, cooperation and coordination of such a motley crowd of micro-units. Treating them as a unique ``representative'' firm or household clearly risks throwing the baby with the bathwater. 
As we have learnt from statistical physics, understanding and characterizing such emergent properties can be difficult. Because of feedback loops of different signs, heterogeneities and non-linearities, the macro-properties are often hard to anticipate. In particular, these situations generically lead to a very large number of possible equilibria, or even the lack thereof.

Spin-glasses and other disordered systems give a concrete example of such difficulties. In order to tackle these complex situations, new theoretical and numerical tools have been invented in the last 50 years, including of course the replica method and replica symmetry breaking, and the cavity method, both static and dynamic. In this chapter we review the application of such ideas and methods in economics and social sciences. Of particular interest are the proliferation (and fragility) of equilibria, the analogue of satisfiability phase transitions in games and random economies, and condensation (or concentration) effects in opinion, wealth, etc. 

\section{Game Theory}
\label{sec:game_theory}

Can we expect cooperation in a population of selfish interacting individuals? Will a particular technology be adopted? How will firms behave when competing in the same sector? What is the optimal way of choosing a route to destination for a taxi driver in a crowded city? All these questions involve phenomena whose outcomes depend on how humans interact with each other. 

Game theory is a general, idealized framework that addresses these questions under the assumption that each individual behaves rationally in order to maximize his/her utility. Because such an assumption is quite unreasonable, the predictions of game theory -- the so-called Nash equilibria -- are often incorrect when compared to real life outcomes. Yet, they provide a useful benchmark for understanding the richness that may result from human interaction. This is particularly true in ``large games'', when statistical behavior with its regularities is expected to set in. 

There are two dimensions with respect to which a game can be ``large'': one is when the number of available strategies\footnote{A strategy is a possible course of action in the game.} for each individual player becomes large, the other is when the number of players itself becomes large. The generic outcome in ``large games'' is that the number of possible Nash equilibria may become very large, i.e. exponential in the number of players or strategies, much like the number of equilibrium states in spin-glasses \cite{mezard1987}. 

Take for example a simple game: $N$ individuals have to decide whether to take one of two routes of the same length, to go from A to B. If their travel takes a time that decreases with the number of people who took their same choice, it is intuitive that the optimal outcome is one where the population will split exactly in two equal parts over the two choices. 
There are $\binom{N}{N/2}\sim e^{N\log 2}$ arrangements of this type. 

When heterogeneity is taken into account, either because strategies are different or because the agents are different, the analysis of large games reveals a rich phenomenology with multiple equilibria, phase transitions and complex behavior, that calls for methods of statistical physics of disordered systems. 

Two-player games, each with many different strategies were considered in Refs.~\cite{rieger1989solvable,opper1992phase,galla_farmer}, within an evolutionary setting\footnote{The equilibria of a game can be thought of as the equilibria of an ecosystem of species each of which plays one of the strategies, against randomly chosen opponents in a well mixed population.}, and then by Berg and Engel~\cite{berg1998matrix,berg2000statistical}, who analyzed games of two players, each of which has $N$ possible strategies. These are called bi-matrix games, because the payoffs to the players depending on the strategy choices of players can be encoded in two $N\times N$ matrices. When these matrices are random, Berg and Engel~\cite{berg1998matrix,berg2000statistical} show that the number of Nash equilibria $\mathcal{N}\sim e^{NS}$ is exponentially large in $N$ and that the entropy $S$ can be computed using the replica method. Given such degeneracy, the game's outcome can hardly be predicted. Yet, typical Nash equilibria share characteristic properties that provide statistical predictions on the outcomes that we may expect in these settings\footnote{For example, it is possible to estimate the number of strategies that will be actually played by the two players, among the $N$ possible ones.}. 

The replica and cavity method has also allowed one to shed light on the typical properties of games of heterogeneous interacting agents, in the limit where the number $N$ of agents diverge. The Minority Game (MG)~\cite{challet2004minority} is probably the prototypical model in this class. It describes the competition of many agents with different strategies on a set of resources. It was introduced as a simplification of the El Farol bar problem, originally proposed by Brian W. Arthur~\cite{arthur1994inductive} as a critique of the deductive rationality approach assumed in game theory. The MG will be discussed in more details in Section~\ref{sec:financialmkts}. 
In brief, it extends the example of choosing the least congested route discussed above, to the case where the origin and destination of each agent differ and the available routes for each of them intersects in complex ways with the one of others. In these situations, the MG assumes that each agent adapts to his/her ``environment'', neglecting the fact that they also contribute to it (i.e. to traffic congestion). Hence the MG is not really a game, because agents do not behave strategically. As we shall see in Section~\ref{sec:financialmkts}, this makes the equilibrium of the game unique, i.e. replica symmetric~\cite{challet2004minority}. 
When instead agents behave strategically (as in game theory) taking into account their systemic impact, replica symmetry is broken, and the game features an exponential number of Nash equilibria~\cite{de2001replica}. 

Players' heterogeneity may also result from the fact that each of them interacts with a different sub-set of players. In these {\em network games}~\cite{jackson2015games}, each player sits on the nodes of a network and interacts only with his/her neighbors\footnote{Network games also include problems where the network itself -- i.e. whom to interact with -- can be chosen strategically by the players~\cite{jackson2015games}.}. The ensemble of games where the network is chosen at random can be studied with cavity methods borrowed from the theory of disordered systems. Consider, for example, a ``local public good'' setting where each player can exploit a resource either by buying it, at some cost, or by free riding on the resources bought by their neighbors. Given a network, who should invest in buying the resource? 
Dall'Asta {\em et al.}~\cite{dall2011optimal} show that finding all Nash equilibria of this game is equivalent to finding all maximal independent sets of the network, which is an NP-hard problem. Still, the number of Nash equilibria and their typical properties in the limit of infinite networks can be computed using belief propagation approaches. Dall'Asta {\em et al.}~\cite{dall2012collaboration} use similar methods in order to understand under what conditions players in a social network can sustain cooperation, in a repeated prisoners' dilemma game.

\section{Choice under Social Influence \& ``Bandwagon'' Goods}
\label{sec:rfim}

As mentioned above, the success of the replica and cavity methods in dealing with complex physical systems has triggered a general interest in the physics community for the study of many complex systems outside the traditional domain of physics, like socio-economic systems (whether or not these require to rely on such advanced statistical physics techniques). On the methodological side, the use of well controlled agent-based simulations, further discussed in Section \ref{sec:firms} below, is one example of physicists contributions, as such simulations often reveal collective, emergent phenomena typically encountered in statistical mechanics. 

At the conceptual level, the importance of multiple equilibria and the difference between quenched and annealed disorder can be quite subtle (see e.g. \cite{mezard1987,benarous}) and not always fully understood in the economics literature. One reason comes from the standard view point in theoretical economy based on game theory, introduced in the previous section. At each instant of time, the `true' underlying dynamics is supposed to be known to the players who compute the Nash equilibria, and then simultaneously play Nash, anticipating that all the other agents will do the same. Even with rational agents, such anticipations are not possible whenever there are multiple equilibria.
Although, as reviewed above, many games turn out to have a large number of equilibria, the classical view in economics is either that models should be defined such that there will be a single equilibrium, or that some postulated dynamics {\em justifies} the selection of one of the equilibria -- but such dynamics usually does not correspond to any actual agents' dynamics! 

`Heterodox' approaches in economics are more in line with the physicist approach, modeling the dynamics of agents with limited rationality~\cite{kirman2010complex} and plausible learning rules~\cite{sato2003coupled, challet2004minority}. The analysis of any such dynamics requires to specify the time scale over which the interactions or other parameters change with time. 

One example of interest is the general problem of the collective behavior of agents making a binary choice under social influence, such as buying or not buying a fashion good, sorting or not sorting the waste, joining or not a riot, etc. Each agent $i$ has their idiosyncratic willingness to adopt (or to buy), $h_i$. This willingness is increased if others do the same (a case of "positive externality"), so that if the price (or cost) of the active decision is $P$, agent $i$ wants to make the decision iff $$h_i-P + J \eta > 0 \ ,$$ where $J>0$ is the strength of the social influence and $\eta$ the current fraction of adopters. 

When the  idiosyncratic willingness $h_i$ are random but time independent, this specification is equivalent to the mean-field Random Field Ising model (RFIM). Such model is a particular instance of the {\em random utility model}\cite{manski77} in theoretical economy and is also equivalent to the `dying seminar' model of the social scientist T. C. Schelling~\cite{schelling78}. With an approach more similar to the one of physicists than that of economists, Schelling correctly inferred the generic existence of multiple equilibria and hysteresis, leading to the notion of critical mass (or tipping point). Ising (or ``Markov Random Fields" in the mathematical terminology) and RFIM type models have been used to model socio-economic systems by mathematicians~\cite{Follmer74}, economists~\cite{BrockDurlauf01,Blume95,GlaeserScheinkman}, physicists~\cite{weidlich71,GalamGefenShapir82,WeiSta,NaPhGoVa06}, to mention but a few: see e.g ~\cite{bouchaud2013crises} for a review and further references, in particular in the context of multiple choices~\cite{borghesi}.  

The statistical physics approach leads to the detailed study of the phase diagram in the parameter space (mean willingness to adopt $\langle h \rangle$ and strength of social influence~$J$)~\cite{NaPhGoVa06}. Considering the vicinity of the critical point, one can predict a scaling law in the case of continuous change of (collective) behaviour~\cite{michard-bouchaud}: the height of the variation peak should scale with the width $w$ as $\sim w^{-2/3}$. This prediction appears to be in good agreement with empirical data on cell phones adoption and birth rates evolution. This is a strong support to the relevance of this type of modeling, since it is hard to think of an alternative argument that would lead to such anomalous scaling. 

In an economic context, considering that the seller is a profit-maximizer leads to the appearance of systemic risk. Indeed, the price of the good that would maximize the seller's profit is generically very close to the value at which the demand, corresponding to a large number of buyers, disappears abruptly. At this critical price value, one indeed crosses the line separating a regime with two Nash equilibria (the `fashion good' equilibrium with a high demand and the low-demand equilibrium) and a regime with a single Nash equilibrium, for which only the rare agents with a very high idiosyncratic willingness to pay buy the good~\cite{GNPV_entanglement}. Hence, for prices slightly higher than the optimal price, demand falls precipitously! 

This can be called a ``cliff-edge'' maximization situation: optimization can be tantamount to fragility. As discussed in  section \ref{sec:firms}, this situation is common in complex systems, which tend to spontaneously stabilize close to a point where the system becomes unstable (on this topic, see also \cite{wyart_marginal}).

\section{Opinion dynamics}
\label{sec:opinion}
There is a large literature on the modeling of opinion dynamics. The boundaries of this research domain are ill-defined: many models of social behavior may be interpreted as models of opinion dynamics, including the binary choice RFIM discussed in the previous Section~\ref{sec:rfim}. Another example is given by models of language change resulting from interactions between locutors, which can be seen as describing a dynamics of opinions on the meaning of words~\cite{Castellano-Fortunato-Loreto}.  

In this section, we restrict to models more specifically related to opinion dynamics. 
The main opinion dynamics models (with their variants) are the Voter model (see~\cite{Castellano-Fortunato-Loreto} and references therein), the Hegselmann and Krause model~\cite{Hegselmann-Krause}, the Deffuant bounded-interaction model~\cite{DeffuantNeauAmblardWeisbuch2000}, the Sznajd model~\cite{Sznajd-Weron_Sznajd}, and the Galam consensus model~\cite{GalamGefenShapir82} based on the RFIM (see Section~\ref{sec:rfim}). All these models explore the hypothesis that opinion may change due to interactions with others, with an imitation behavior -- a search for consensus. For general reviews of opinion dynamics model inspired by statistical mechanics, see~\cite{Castellano-Fortunato-Loreto,Jedrzejewski_Sznajd-Weron_2019}. Although not fully developed, the use of the cavity method to understand how opinions or rumors spread on a network like epidemics (see e.g. \cite{moreno,newman,anand2013epidemics, kuhn,torino1,torino2}) is surely an interesting path. Here we add a note on the less often mentioned Seceder model~\cite{Dittrich_etal_2000} exploring a different hypothesis, the case where every agent wants to imitate those who are not like everyone else.

Dittrich {\em et al.}~\cite{Dittrich_etal_2000} have introduced a model of agents trying to adopt opinions/behaviors  different from the ones of others.  In the context of genetic evolution (genes being analogous to opinions), the model explores the outcome of giving an advantage to individuals sufficiently different from the others (see also ~\cite{courson}). The authors show that clusters can emerge from a dynamics with such rule. 
Soulier and Halpin-Healy~\cite{SoulierHalpin-Healy2003} have considered a simple variant with such dynamics ``pitting conformity against dissent". Opinions are described by vectors of continuous variables in $d$ dimension. At each time step, one randomly selects an agent (the voter). Then one picks at random a set of $m$ individuals, the polling group. The individual with opinions most distant from the mean opinion of the polling group is selected. Then the voter adopts, with small variations in opinion values, the opinions of this group-dissident agent.

The model behavior changes abruptly with the group size at small values of $m$. A surprise is that the mean field behavior is reached at $m=4$. In opinion dimension $d=1$, for $m\geq 4$, the population clusters into two groups. The authors consider a variant with discrete opinions, in which the randomly chosen agent adopts exactly the opinions of the distant outlier of the group. They can then write deterministic equations of the replicator type. The analysis of the model leads to a quite remarkable result. If $d$ is equal or larger than $3$, and $m\geq 4$, the dynamics always leads to a condensation of opinions in a space of two dimensions, with the emergence of three clusters in the space of opinions. 

Most of the opinion dynamics model have been introduced on the basis of theoretical motivations. As pointed in~\cite{redner_cras2019}, from the study of many variants one observes non-robust results when rules are changed. Although it is useful to explore the space of possibilities, and interesting to discover a rich variety of model behaviors, this calls for restricting such studies to more empirically based models. 

One should note that most models do not really make the difference between `having an opinion' and `making a decision' -- except for the fact that a decision may be the noisy outcome of the opinion. When considering voting behavior, models do not address issues of strategic behaviors.

Only few works try to test models on data or define data-specific models (see e.g.~\cite{GalesicStein2017}). However, models with rules inspired by the social psychology literature are being considered~\cite{Jedrzejewski_Sznajd-Weron_2019}. This allows one to study the outcome of behavioral rules motivated by experimental findings. Much work remains to be done along such lines. The increasing possibility to access to data and to perform online experiments should trigger studies more directly linked to empirical behavior.


\section{Firm Networks, Ecologies \& Portfolios} 
\label{sec:firms}

There is a renewed interest in models of networks of interacting firms as a possible framework that accounts for excess fluctuations in economic systems, arising from ``contagion'' or default cascades that propagate along the supply chains (see \cite{carvalho} for a recent review). Similar ideas also exist in the context of bank networks \cite{bank_networks, gai2011complexity} or overlapping portfolios \cite{caccioli_farmer} to account for banking crises and deleveraging spirals in financial markets.

The economic behavior of firm networks depends on the {\it {production function}} that models how input goods and labor are transformed into a certain product. A classic production function that expresses {\it non-substitutability} of inputs is the Leontief model, that reads:
\be 
\pi_i = z_i \min_j \left\{ \frac{Q_{ij}}{J_{ij}} \right\} \ ,
\ee 
where $\pi_i$ is the amount of goods firm $i$ produces, $z_i$ is the so-called productivity of firm~$i$, $Q_{ij}$ is the amount of goods $j$ available to $i$ and $J_{ij}$ are similar to ``stoichiometric coefficients'' in chemistry, measuring how many units of $j$ are needed to build $i$ (when $z_i=1$). Conventionally, labor corresponds to good $j=0$. 

Once the production function is specified, the equilibrium state of the economy is obtained by imposing (a) that firms attempt to maximize their profit and (b) that markets clear, i.e. that everything that is produced is consumed. These two conditions translate into enough equations to determine equilibrium prices $p_i$ and productions $\pi_i$. For example, for a Leontief production 
function the equations for the vector of prices $\vec p$ (given in units of the cost of labor) read \cite{moran_bouchaud}
\be \label{eq_eq}
\mathbb{M} \, \vec p = \vec V \ , \qquad \mathbb{M}_{ij} = z_i \delta_{ij} - J_{ij} \ ,
\ee 
where $V_i = J_{i0}$.

The problem with such a set of equations is that the solution $\vec p$ is not necessarily a positive vector \cite{hawkins_simon}. In other words, some constraints must be fulfilled by the $z_i$'s and the $J_{ij}$'s for the economy to be viable.\footnote{When all $J_{ij}$ are positive, the matrix $\mathbb{M}$ is an ``M-matrix'' and the constraints boil down to imposing that all eigenvalues of $\mathbb{M}$ have a positive real part.} If such constraints are not satisfied, some firms (the least productive ones) must necessarily be removed for the economy to become viable. So, much as counting the number of equilibrium states in spin glasses, one can ask the following question: for given productivities $z_i$ and interaction matrix $J_{ij}$, how many viable economic equilibrium states are possible as a function of the number of firms $N$? And for a generic equilibrium, what is the eigenvalue spectrum of $\mathbb{M}$, which determines the dynamical stability of such an equilibrium?      

Interestingly, the very same questions arise in the context of the Lotka-Volterra description of complex ecological networks. Denoting now as $p_i$ the population size of specie $i$, the equilibrium states of the Lotka-Volterra equation are such that
\be 
p_i \left(V_i - z_i p_i + \sum_{j\neq i} J_{ij} p_j \right) = 0 \ ,
\ee 
where now $V_i$ describes the fitness of specie $i$, $z_i$ is a saturation (self-interaction) term, and $J_{ij}$ models the interactions between species: $J_{ij} > 0$ means that the presence of specie $j$ favors the growth of specie $i$, whereas $J_{ij} < 0$ means that the presence of specie $j$ hampers the growth of specie $i$. Ecological equilibria are thus such that either $p_i = 0$ (i.e. specie $i$ is extinct) or, for all remaining species, Eq.~\eqref{eq_eq} leads to a positive solution. When $J_{ij}$ is a symmetric matrix, the Lotka-Volterra model can be mapped onto a spin-glass problem \cite{biroli_bunin, altieri_biroli}. 

For independent random symmetric $J_{ij}$, there exists a replica symmetry broken phase, corresponding to a proliferation of possible ecological equilibria. 
These equilibria are furthermore found to be {\it marginally stable}, in the sense that the eigenvalue spectrum of $\mathbb{M}$ touches zero, meaning that such equilibria are extremely fragile to perturbations, for example small changes in the interaction matrix $J_{ij}$ \cite{biroli_bunin}. In this phase, evolution naturally leads to a self-organized critical state, where the stability criterion proposed by R. May in his famous paper is exactly saturated \cite{may1972} (see also \cite{fyodorov_may1, fyodorov_may2} for further developments). One can speculate that economies, too, spontaneously evolve towards a marginally stable state, for which small external shocks may lead to anomalously high volatility \cite{moran_bouchaud}. 

Yet another, completely different setting where the very same mathematical discussion arises is portfolio construction with constraints. Markowitz' celebrated optimal portfolio (also known as mean variance optimization) states that the weight $p_i$ of asset~$i$ should be chosen as the solution of Eq.~\eqref{eq_eq}, where now $\mathbb{M}$ is the covariance matrix, measuring how the returns of asset $i$ and asset $j$ are correlated, and $\vec V$ is the vector of predicted gains for each asset. Positive weights correspond to long positions, whereas negative weights indicate that the asset manager should go short the corresponding asset. 

But more often than not, asset managers cannot take short positions. In other words, their portfolios must satisfy a positivity constraint, $p_i \geq 0$, $\forall i$, much like prices in the firm network model and population sizes in the Lotka-Volterra model. So we are again back to the same ``spin-glass'' type problem \cite{galluccio, mezard_ciliberti, garnier_brun, farkas1, PhysRevE.101.062119, farkas2}: what is the number of positive solutions of Eq.~\eqref{eq_eq} as a function of the number of assets and the parameters of the problem? A fully soluble case is when there is a unique risk factor that correlates the returns of different assets \cite{garnier_brun}. This corresponds to $\mathbb{M}_{ij}=z_i \delta_{ij} + \beta_i \beta_j (1-\delta_{ij})$, where $\beta_i$ is the exposure of asset $i$ to the common risk factor. One finds that for such a problem, optimal portfolios are typically sparse, and the total number of solutions grows sub-exponentially with the number of assets -- whereas spin-glass problems typically have an exponentially large number of local optimums. But, in common with spin-glass problems, one can find very different, quasi-degenerate optimal portfolios in the presence of non-negative constraints, or other types of non-linear constraints (see also \cite{farkas1, farkas2}). 

In fact, many problems in economics and finance are constrained optimization problems, or constraint satisfaction problems (see e.g. \cite{dhruv1, dhruv2} for a recent example). It is expected that many of these problems share with spin-glasses two important properties, typical of replica symmetry broken systems: a) the existence of a large number of quasi-degenerate solutions, that are nearly equally good in terms of their performance but very far from one another in phase space (e.g. surviving firms, surviving species, assets with non-zero weights in the above examples); b) parameter ``chaos'', i.e. the sensitivity of these solutions to the precise specification of the parameters of the problem \cite{fisher_huse_chaos, bray1987chaotic, rizzo2003chaos}. In other words, the optimal solution for one choice of parameters can become suboptimal, or even disappear, when these parameters are only 
slightly changed (in the limit of a large number of degrees of freedom $N$: number of firms, number of species, number of assets). 

Explosion of the number of optimal solutions and parameter chaos raise many difficulties in the modelling of complex systems. The classical approach in terms of probabilities is doomed by non-ergodicity, and the need to think in terms of probabilities of probabilities, like in spin-glasses \cite{parisi_complexity}. Such difficulties, that one could coin as ``radical complexity'' \cite{radical_complexity} with a nod to Keynes' ``radical uncertainty'', should lead to significant rekindling of the way complex socio-economic problems are addressed. Two particularly interesting directions are (a) minority games and complex game theory and their relation with spin-glasses, see \cite{challet2004minority, galla_farmer, berg2000statistical} and section \ref{sec:game_theory}; (b) agent based modelling and scenario identification \cite{tipping_points, dhruv_covid, radical_complexity}.   

\section{Replica Method for Financial Markets and Large Random Economies}
\label{sec:financialmkts}

The mutual attraction between finance and physics has diverse roots. One is to be found in the financial industry's thirst for analysts with a quantitative training. On the other side, the increasing availability of financial data attracted the curiosity of physicists interested in understanding the non-trivial statistical features of market prices, which suggest analogies with fluid turbulence, avalanches, earthquakes and other phenomena in natural sciences. 

On the theoretical side, the prevailing neo-classical paradigm relies on the pillars of the No-Arbitrage Hypothesis and the Efficient Market Hypothesis (EMH) ~\cite{samuelson1965proof,fama1970efficient} -- according to which price behavior can not be predicted and hence no excess gain can be extracted from speculative trading. This provides a theoretical foundation for asset pricing, but it was soon realized that these assumptions lead to conceptual inconsistencies~\cite{grossman1980impossibility} and that observed market behavior is incompatible with the neo-classical assumption of traders with rational expectations. (There is an enormous literature on this last point, for a physicist viewpoint and many references, see~\cite[Chapter 20]{TQP}.) 

On the other hand, early simulations of agent based models~\cite{palmer1994artificial} have shown that simple models with way less sophisticated traders could reproduce the main stylized facts observed in real financial markets. This spawned a line of research (see~\cite{farmer2000physicists} for an early review, and~\cite{pietronero} for a more recent one) aimed at understanding financial markets within simple models of interacting adaptive agents, that could explain the observed stylized facts and be amenable to theoretical analysis (see e.g.~\cite{bak1997price,brock1998heterogeneous,cont2000herd} for some early examples). 

\subsection{The Minority Game}

The Minority Game~\cite{challet2004minority} was introduced~\cite{challet1997emergence} as one such attempts. It depicts the interaction between financial traders as follows: in order to out-compete other traders, a trader needs to anticipate when most of them will buy, so that the trader can sell at a high price (or vice-versa). Each trader then aims at being in the minority group of either sellers or buyers. Note that this interaction promotes the diversification of strategies across traders. A strategy prescribes whether to buy or to sell, depending on the pattern of the $m$ most recent signs of the price fluctuations. Therefore each strategy is a table of $P=2^m$ numbers, and each trader is assigned few (say two) randomly drawn strategies. Each trader evaluates the performance of his/her strategies in the course of time and then plays the best one -- i.e. the one that would have placed him/her most often in the minority --- at each time step. 

The stationary state of the MG can be fully analyzed with techniques coming from the statistical mechanics of disordered systems~\cite{challet2000statistical,coolen2005mathematical}. The main insight that the MG provides on the behavior of financial markets is that, as the number of traders $N$ increases, the market becomes less and less predictable and, at a critical value $n_c$ of the ratio $n=N/P$, it becomes completely unpredictable, meaning that the expected value of future returns is independent of the past history. The point $n_c$ marks a second order phase transition between a symmetric (information efficient) phase for $n\ge n_c$ and an asymmetric (inefficient) phase. In other words, the MG provides a stylized description of how a market becomes informationally efficient as more and more (diverse) traders join it. The phase transition is also accompanied by critical fluctuations similar to those observed in the statistics of real returns~\cite{challet2003criticality}, which suggests that the observed anomalous fluctuations in financial markets are the other side of the coin of market's information efficiency. The relation between market efficiency and criticality emerges also in other modelling approaches to financial markets, see e.g.~\cite{bouchaud2003fluctuations,TQP}.

The MG proved to be a wellspring of further interesting results. Cavagna {\em et al.}~\cite{cavagna1999thermal} introduced a thermal version of the MG whereby agents chose the strategy they play stochastically. Detailed analysis of the dynamics~\cite{marsili2001continuum} reveals that the ``temperature" introduced at the microscopic scale in this way turns out to play the role of the inverse of a ``temperature" at the collective level. The symmetric phase exhibits a peculiar type of broken ergodicity where the stationary state ``remembers" the initial conditions, e.g. the prior beliefs of agents. The replica theory of MG provided a playground for addressing several issues, such as the effect of a Tobin tax~\cite{bianconi2009effects} or a prediction of the market impact of meta-orders\footnote{Meta-orders are long sequences of orders in the same direction, i.e. either buy or sell, by the same investor. A key issue is whether meta-order have a permanent effect on prices or not~\cite{toth2011anomalous, TQP}. In the MG the permanent impact can be computed analytically and it is non-zero only in the asymmetric phase, whereas it vanishes when the market is unpredictable.}~\cite{barato2013impact}. 

The mechanism underlying the phase transition in the MG is rather generic. It also describes information aggregation in an asset market with traders with heterogeneous information~\cite{berg2001statistical}, leading to conclusions similar to those discussed above. 

\subsection{Large Random Economies}

Indeed, the nature of such a phase transition is similar to that occurring in large random systems of linear equations with non-negative constraints (see e.g.~\cite{farkas1,PhysRevE.101.062119}), that can describe ecologies, firm networks and economies, see Section \ref{sec:firms}. The financial industry as a whole can be considered as a large random economy and therefore the statistical mechanics analysis, thanks to the replica method, can shed light on the consequences of the expansion of the repertoire of financial instruments, such as the one we have witnessed since the nineties. The neo-classical lore maintains that the more financial instruments consumers have at their disposal, the better they can hedge their risks. The ideal situation is that of {\em complete markets}, when the repertoire of financial instruments is so large that risk can be eliminated altogether, as in Black-Scholes theory of option pricing~\cite{hull2006options}. 

The replica method however reveals that the quest for market completeness can lead to financial instability~\cite{marsili2014complexity}. The volumes of interbank trading necessary to hedge all financial instruments, as well as the susceptibility of the equilibrium to exogenous shocks, diverge as the financial sector approaches the ideal limit of complete markets even in an ideal model. Similar conclusions were derived within a different modelling approach~\cite{brock2009more}, suggesting that the same may apply to real markets. As already mentioned above, ``efficiency'' and ``optimality'' can lead complex systems to the brink of instability. 

\begin{figure}[t]
\centering
\includegraphics[width=0.9\textwidth]{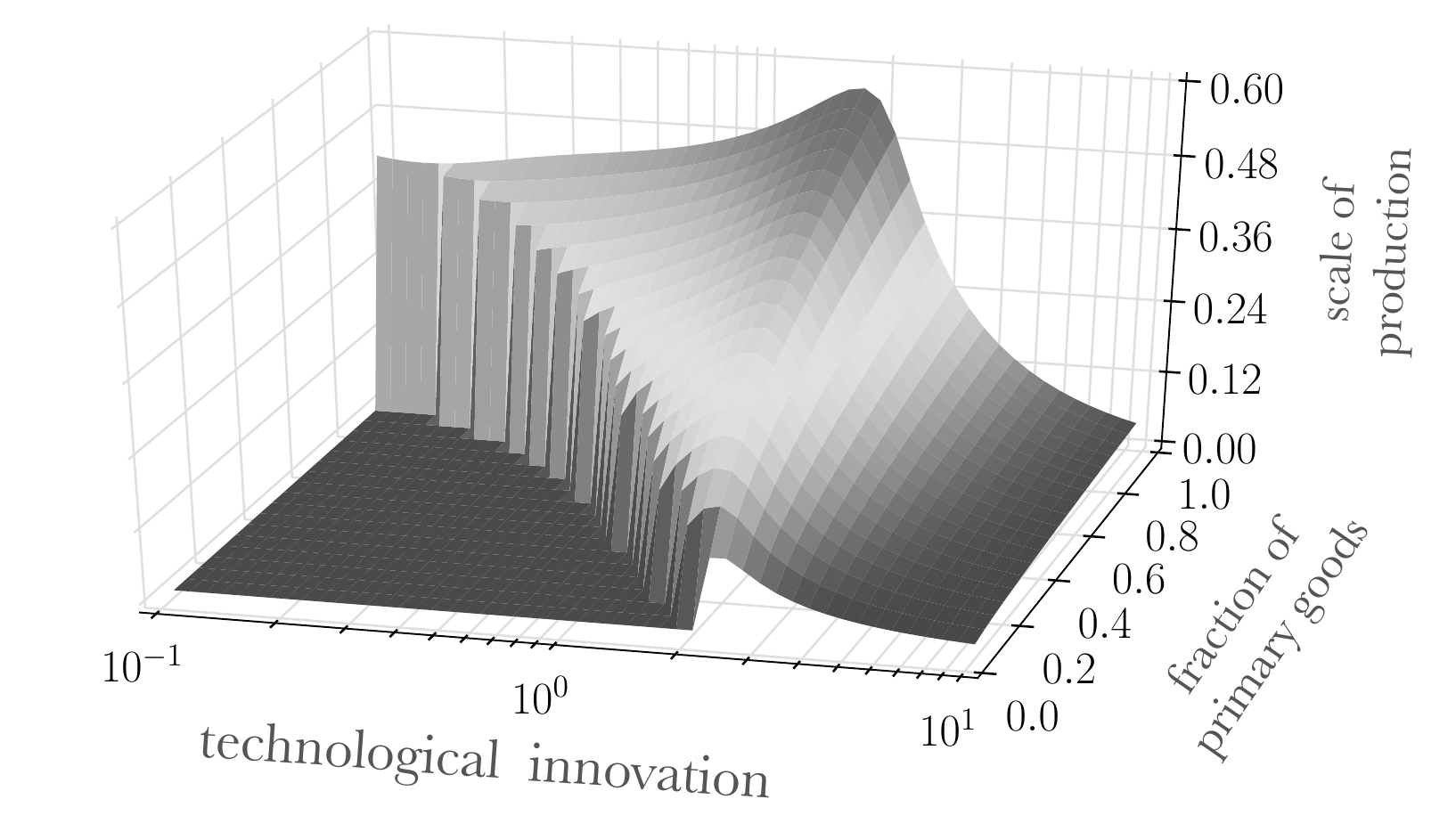}
\caption{Phase diagram of the General Equilibrium of a large random economy, from~\cite{bardoscia2017statistical}. The technological innovation axis is the ratio between the number of technologies and the number of goods. 
The introduction of new technologies amounts to a shift on the right along this axis. The introduction of new goods reduces both this coordinate and the fraction of primary goods, so it corresponds to shifts towards the origin.}
\label{fig:geneq}
\end{figure}

The effects of technological innovation in a large random production economy can be analyzed in a similar way. In this economy, firms produce final goods for household consumption from either primary goods or inputs produced by other firms, through a linear transformation. Each firm's technology is then a random vector in the space of commodities. In this standard set-up~\cite{lancaster2012mathematical}, firms maximize profit, consumers buy and consume final goods maximizing their utility and market prices are fixed by market clearing. Bardoscia {\em et al.}~\cite{bardoscia2017statistical} show that a statistical mechanics analysis of the way the equilibrium of such a large random economy changes (as more and more technologies are ``invented'') has suggestive similarities with the industrial development we have witnessed so far. As long as the number of technologies does not exceed a given threshold\footnote{Which is proportional to the number of goods. In the statistical mechanics analysis both the number of goods and the number of technologies diverge, with a fixed ratio.} consumers can rely only on those final goods that are also primary goods, because no firm can operate. When the number of technologies exceeds this threshold, an ``industrial revolution" takes place as a sharp phase transition. Beyond this point, the introduction of new successful\footnote{A new (random) technology is adopted only if it generated profits at the current prices. It is not adopted if the profit is negative.} technologies increases the scale of production of already existing technologies. This suggests that firms have incentives to incorporate all the technologies needed to produce final goods, as well as to carry out in-house research and development. This is reminiscent of the early stages of industrialization, which was characterized by vertically integrated firms~\cite{langlois2003vanishing}, where all intermediate production processes were managed within the firm. As technological innovation proceeds, the economy crosses over to a regime characterized by a saturated technological repertoire. Beyond this second transition, that occurs close to the maximum in figure~\ref{fig:geneq}, the introduction of further technologies is disruptive: most of the technological innovations are not viable and the few that are successful displace other technologies when they are introduced, reducing their scales of production. In this regime, the growth of the economy is not driven by the introduction of new technologies but by the introduction of new goods. This is evocative of the expansion of markets for intermediate goods, those who are neither primary nor final goods (e.g. parts of a final good, like electronic components) and a parallel outsourcing of segments of the  processes involved in the production of complex goods. These are both processes that advanced capitalist countries have experienced~\cite{langlois2003vanishing}.

The prediction of the stylized picture provided by this model is that, on one side, pursuing economic growth, advanced economies are expected to converge to the critical point separating the two regimes, on the other that the expansion of intermediate goods markets and outsourcing may lead the economy to a collapse, crossing the ``industrial revolution" phase transition point.

\section{Intermittency \& Condensation Phenomena}

\subsection{Sums of Exponentials \& RSB}

In many cases of interest in physics, but also in finance, social sciences and economics, one has to deal with sums of exponentials of random variables. In statistical physics, the {\it partition function} is defined as the sum over all configurations of the Boltzmann-Gibbs weight, which is itself the exponential of (minus) the energy of that configuration divided by temperature. When these energies are random variables, one is confronted to a sum of exponentials of random variables -- this is Derrida's Random Energy Model (REM), which is well-known to be characterised by a 1-RSB glassy phase at low temperatures \cite{derrida_REM, gross_mezard, bouchaud_mezard_extremes}. Population growth or survival \cite{benarous}, city growth \cite{gabaix} or wealth growth \cite{bouchaud_mezard_wealth} are other examples where such sums of exponentials naturally appear. 

So let us consider, generically, 
\be
Z_N = \sum_{i=1}^N z_i \ , \qquad z_i := e^{t \xi_i} \ ,
\ee 
where $t$ is a positive parameter and $\xi_i$ are {\it i.i.d.} random variables, that we can always choose to be of zero mean and unit standard deviation. We also assume that the right tail of the distribution of $\xi$'s decays as 
\be \label{eq_rho_asympt}
\rho(\xi) \sim_{\xi \to \infty} C e^{-A \xi^b},
\ee 
where $A,C$ are positive parameters and $b > 1$ (corresponding to super-exponential decay). When $b=1$ the distribution of $z$'s has a power-law tail for large $z$, $z^{-1-\mu}$, with a tail exponent $\mu = A/t$, such that $n$-th moments with $n \geq \mu$ are divergent. When $b > 1$, on the other hand, all moments $\mathbb{E}[e^{n t \xi}]$ exist and are finite when $t \geq 0$. Hence, formally, the Central Limit Theorem (CLT) applies and, for large $N$, $Z_N$ should converge to a Gaussian distribution. In particular, the Law of Large Numbers suggests that for large~$N$ 
\be \label{eq:LLN_exp}
Z_N \approx N \mathbb{E}[e^{t\xi}] \ .
\ee 
However, consider the case where $N$ is large but finite and take $t$ to infinity first. Then, the whole sum $Z_N$ will be concentrated in its largest term:
\be 
Z_N \approx_{t \to \infty} e^{t \xi^\star} \ ,
\ee 
where $\xi^\star = \max_{i=1, \dots, N} \xi_i$. We thus see that depending on whether $N$ is taken to infinity at fixed $t$, or $t$ is taken to infinity at fixed $N$, the result is markedly different. 

More generally, one finds that for a given, large value of $N$, the relevant tail of the distribution of $z$ is again a power-law, but with an $N$-dependent tail exponent:
\be \label{eq_mu_exp}
\mu = \frac{b A^{1/b}}{t} \left( \log N \right)^{1 - \frac{1}{b}}.
\ee 
We are now in position to state the following result, which generalizes the 1-RSB transition of the REM. Suppose that both $N$ and $t$ go to infinity, with $\mu$ (as given by Eq.~\eqref{eq_mu_exp}) fixed. Then, depending on the value of $\mu$, the sum of exponentials of random variables, $Z_N=\sum_i \exp(t \xi_i)$, obeys either the standard CLT or the generalized (L\'evy) CLT \cite{derrida_3levels,benarous,bovier_kurkova}:
\begin{itemize}
    \item $\mu > 2$: $Z_N$ converges towards a Gaussian random variable of mean $N \mathbb{E}[e^{t\xi}]$ and variance $N (\mathbb{E}[e^{2t\xi}] - \mathbb{E}[e^{t\xi}]^2)$.  
    \item $1 < \mu < 2$: $(Z_N - N \mathbb{E}[e^{t\xi}])/N^{1/\mu}=u$ converges towards a L\'evy-stable random variable, with $P(u) = L_{\mu, \beta=1}(u)$, the totally asymmetric L\'evy distribution of index $\mu$.
    \item $\mu < 1$: $Z_N/N^{1/\mu}=u$ converges towards a L\'evy-stable random variable, with, again, $P(u) = L_{\mu, \beta=1}(u)$, the totally asymmetric L\'evy distribution of index $\mu$. 
\end{itemize}
Hence, for $\mu > 2$, the {\sc{LLN}} result holds (see Eq.~\eqref{eq:LLN_exp}), but completely falls apart when $\mu < 1$.

Take for simplicity $\rho(\xi)=\exp(-\xi^2/2 \sigma^2)/\sqrt{2 \pi \sigma^2}$. In this case, $A=1/(2 \sigma^2)$ and $b=2$, so that $\mu = \sqrt{2\log N}/(\sigma t)$. What we learn from the previous discussion is that for small enough $t$, $Z_N$ is Gaussian and the inverse participation ratio (called the Herfindahl index in social sciences), defined as:\footnote{This quantity is often noted $Y_2$ in the spin-glass literature, see e.g. \cite{derrida_3levels}.}
\be
\mathcal{H} = \frac{\sum_{i=1}^N z_i^2}{Z_N^2},
\ee 
is of order $N^{-1}$, i.e. the whole sum is spread out over all elements. As $t$ increases, the Herfindahl index increases and $Z_N$ becomes more and more concentrated in a few terms. When $t > t_{\text{c}} =\sqrt{2\log N}/\sigma$, the Herfindahl index becomes of order unity, even as $N \to \infty$. This corresponds to a genuine ``condensation'' or ``localization'' transition, with many relevant applications, in particular in the context of the glass transition \cite{biroli_bouchaud}.    

As a vivid illustration, imagine a portfolio composed of many different assets, each with a different rate of return $r_i$. At time zero, the total capital $K_0$ is invested uniformly across all these $N$ assets. After time $t$, the capital has accrued and is given by 
\be 
K(t) = \frac{K_0}{N} \sum_{i=1}^N e^{t r_i}.
\ee 
The above analysis tells us that there exists a critical time $t_{\text{c}}$ beyond which the initially diversified capital becomes concentrated among a small subset of assets. Furthermore, at large $t$, the growth rate of $K(t)$ is given by $r^\star= \max \{ r_i \}$. A way to avoid such a condensation phenomenon is to periodically rebalance the portfolio, redistributing the profit and losses of the portfolio among all assets. But if the redistribution graph is sparse, or low dimensional, condensation still takes place if the redistribution rate is small enough \cite{bouchaud_mezard_wealth}. One can similarly consider a multiplicative growth model for firms, wealth, cities, biological species, etc. The model can also be used in the context of the famous ``exploration-exploitation'' tradeoff \cite{gueudre_et_al}.

Note that these random growth with redistribution problems map onto the ``Directed Polymer'' problem -- a ``baby spin-glass'' problem where disorder and interactions compete. The case of directed polymers on a tree-like graph, introduced by Derrida \& Spohn \cite{derrida_spohn}, can in fact be solved using either replicas, or propagating front methods (see also \cite{carpentier_ledoussal, gueudre_et_al}). The ``condensed'' phase corresponds to the pinned, glassy phase of the directed polymer. The Derrida-Spohn model has also strong connections with the multifractal model of financial price series, as discussed below. 

\subsection{Multiscaling, Intermittency \& RSB}

The scaling behavior of the different moments of a random variable brings us to the topic of {\it multiscaling}, which is an important feature of intermittent systems, like turbulent flows where it was first discovered (for a review, see \cite{frisch1995turbulence}). Consider a fluctuating time series $x(t)$, for example the velocity in a turbulent flow, the (log-)price of a stock, or the output level of an economy, etc. 

A natural question to ask is: how much does $x(t)$ varies between two instant of times? One usually first ``detrends'' the time series by removing a (generalized) drift, defined as:
\be 
m_1(\tau) := \langle x(t+\tau) - x(t) \rangle_T,
\ee 
where $\langle \dots \rangle_T$ denotes an empirical sliding average over a total interval of size $T$. 

Assuming stationarity in time, the {\it fluctuation} around the trend is often characterized by the variance of the de-trended increments of $x(t)$, i.e.
\be 
\sigma^2(\tau) := \langle \left(x(t+\tau) - x(t) - m_1(\tau)\right)^2 \rangle_T.
\ee 
The simplest example of a random time series is the Brownian motion. Once detrended, the increments $\Delta= x(t+\tau)-x(t)$ are Gaussian random variables with zero mean and variance $\sigma^2(\tau) = \Sigma^2 \tau$. Hence all higher moments can be computed and expressed in terms of $\sigma(\tau)$, as $m_q(\tau)= C_q \, \sigma^q(\tau)$, where $C_q$ are the moments of the standard normal distribution. In other words, the moment of order $q$ simply scales as the $q$-th power of the standard deviation. All moments thus give the same characterization of the time evolution of the fluctuations of $x(t)$. One speaks of ``monoscaling'' in such a case.

This is however not the only possibility. An example coming from the Burgers equation, and again deeply related to 1-RSB and pinning problems, is the following \cite{bouchaud_mezard_burgers, bouchaud_mezard_extremes}. Consider a time series made of upward ``ramps'' of constant slope $S$, separated by downward ``shocks'', i.e. discontinuities of amplitude $-\Delta_0$ appearing at a Poisson rate $\lambda$. We choose the parameters $S$, $\Delta_0$ and $\lambda$ in such a way that our time series has no long term bias. The computation of the different moments $m_q(\tau)$ is easy in the limit where $\lambda \tau \ll 1$, i.e. when the probability $p(\tau) \approx \lambda \tau$ for a shock to exist between $t$ and $t+\tau$ is very small. The scaling behaviour of $m_q(\tau)$ for $\lambda \tau \ll 1$ is found to be very different depending on whether $q$ is $\leq 1$ or $\geq 1$ \cite{bec_burgers}:
\begin{equation}
\begin{cases}
m_q(\tau) \sim  (\lambda \tau)^q  \sim \sigma^{2q}(\tau) \ , \qquad & \forall q \leq 1\ , \\
m_q(\tau) \sim  \lambda \tau \sim \sigma^2(\tau) \ , \qquad & \forall q \geq 1 \ .
\end{cases}
\end{equation}
Note, interestingly, that $m_2(\tau)$ grows like $\tau$ both for the standard Brownian motion and for the ``ramps and shocks'' model, although the underlying processes are clearly very different. This shows that the second moment is totally blind to intermittency effects. 

In order to characterize these intermittency effects, one defines an exponent $\zeta(q)$ from the scaling of the $q$-th moment, as:
\be 
m_q(\tau) \sim \left[\sigma(\tau)\right]^{\zeta(q)}.
\ee 
Any concave deviation away from the monoscaling behaviour $\zeta(q)=q$ is a signature of ``intermittency'', i.e. the concentration of activity (here the variations of $x(t)$) in some particular regions of space and/or time. 

A well studied model of intermittency is the ``multifractal Brownian motion'' (MBM) \cite{bacry_muzy1,bacry_muzy2}. The (detrended) MBM $x(t)$ can be constructed as follows:
\be
{\rm d}x(t) = \Sigma \, e^{\omega(t)} \, {\rm d}W(t),
\ee
where $W(t)$ is a standard Wiener process (or Brownian motion) and $\omega(t)$ is itself a Gaussian random variable with logarithmic correlations, i.e. a very long memory process:\footnote{Here there is a subtlety that we carelessly sweep under the rug: in order to be well defined, the $\log$ function must be regularized for small $\tau$, see e.g. \cite{vargas}.}
\be 
\mathbb{E}\left[\omega(t) \omega(t+\tau)\right] =  \kappa  \max(\log(\tau/T),0) \ , \qquad 0 \leq \kappa < \frac12 \ ,
\ee
where $\kappa$ is the so-called intermittency parameter, and $T$ is a large time scale cut-off, beyond which volatility is uncorrelated. When $\kappa=0$, the process recovers the standard Brownian motion with a constant volatility. 
In financial parlance, $\omega$ is the local log-volatility of the price $x(t)$. One can show that within this model, 
\be 
\sigma^2(\tau) = \Sigma^2 \tau \ ,
\ee 
independently of $\kappa$. In other words, the second moment of $x(t+\tau) - x(t)$ is the same for the standard Brownian motion and for the MBM. For other moments, however, differences appear. In particular, one finds that the 
multi-scaling exponent $\zeta(q)$ is given by \cite{bacry_muzy1}:
\be 
\zeta(q) = q - \kappa q(q-2) \ , \qquad (q \kappa < 1) \ ,
\ee 
i.e. a linear function for $\kappa=0$ (no intermittency) and a concave function for $\kappa >0$ when volatility is fluctuating.  (When $q \kappa \geq 1$, the corresponding moment is infinite: the MBM develops power-law tail increments, with a tail index $\mu=1/\kappa$ \cite{bacry_muzy1}.)

The deep relation with the REM is the following. The MBM is a Brownian motion subordinated to a ``fractal time'' $s$ defined as:
\be 
s^2 := \int_0^t {\rm d}t^\prime \, e^{2 \omega(t^\prime)} \ ,
\ee
which can be seen as the partition function $Z_t$ of a particle in a logarithmically correlated one-dimensional random potential $\omega(\cdot)$ at inverse temperature $\beta=2$. It turns out that this problem has been thoroughly studied \cite{carpentier_ledoussal,fyodorov_bouchaud}, in particular in connection with the Derrida-Spohn Directed Polymer problem and other specific models of glasses, with the Gaussian Free Field in two dimensions and ``multiplicative chaos'' \cite{vargas} and with Random Matrix Theory problems \cite{fyodorov2016fractional}. Physically, we recover the same phenomenology as the Derrida-Spohn model: there is a 1-RSB low temperature phase (corresponding to $\kappa > 1$), although the distribution of low-lying energy states has a tail that is slightly different from that of the Gumbel distribution, which characterizes the pure REM~\cite{carpentier_ledoussal,fyodorov_bouchaud}.  

Note finally that one can generalize such a logarithmically correlated random potential, which corresponds to a 1-RSB REM, to the full replica symmetry broken case, providing an explicit construction of hierarchical Parisi landscapes in finite dimensions \cite{fyodorov_bouchaud2}.

\bibliographystyle{plain}
\bibliography{biblio27}

\end{document}